\documentclass[hyper,letterpaper,11pt]{JHEP3}
\usepackage{amsmath,amssymb,multirow,array}

\title{Hyperscaling-Violation on Probe D-Branes}

\author{Martin Ammon$^a$, Matthias Kaminski$^{b}$, Andreas Karch$^{b}$ \\
$^a$ Department of Physics and Astronomy, University of California, Los Angeles, CA 90095, USA \\
$^b$ Department of Physics, University of Washington, Seattle, WA 98195, USA \\
Email: ammon@physics.ucla.edu, mski@uw.edu, akarch@uw.edu
}

\abstract{
For the field theories dual to D3/D7- and D3/D5-brane systems we find non-relativistic finite
density fixed points exhibiting a violation of
hyperscaling. This violation is measured by the critical exponent
$\theta=1$ while the dynamical critical exponent
is $z=2$. At zero temperature we compute the thermodynamic potentials, the speed of normal sound,
and the speed of zero sound for both these massive D3/D(2$n$+1)-brane systems
near their non-relativistic fixed points. Moreover, we determine the first correction to the free energy for small temperatures yielding the critical exponents $\alpha$ and $\nu$.
}

\begin{document}
\tableofcontents

%%%%%%%%%%%%%%%%%
\section{Introduction}
\label{sec:introduction}
The AdS/CFT correspondence~\cite{Maldacena:1997re,Witten:1998qj,Gubser:1998bc}, and more generally gauge/gravity dualities, provide a powerful tool for studying strongly coupled field theories in states with finite density. Hence gauge/gravity dualities might
be useful in condensed matter physics (see~\cite{Hartnoll:2009sz,Herzog:2009xv,McGreevy:2009xe} and references therein). Of particular interest are compressible states in which by definition the non-zero charge density varies smoothly as a function of the chemical potential. Well-studied examples of such compressible states are solids, superfluids, and Fermi liquids. However, there exist more exotic compressible states in nature. One very interesting but persistently mysterious example of such an exotic compressible state is the 'strange metal' phase in high-$T_c$-superconductors. 

From a theoretical perspective understanding and classifying such exotic compressible states is a great challenge. 
For example, some exotic compressible states show a logarithmic violation of the area law in their entanglement entropy \cite{Ogawa:2011bz}, as well as the existence of hidden Fermi surfaces \cite{Huijse:2011hp,Huijse:2011ef}. The hidden Fermi surfaces are linked to the phenomenon of hyperscaling violation.
The gauge/gravity duality generically provides such compressible phases. In this paper we study two such examples of compressible states near quantum critical phase transitions. In particular we determine the critical exponents including the hyperscaling violation exponent.

Critical exponents traditionally arise in the context of dynamical
critical phenomena \cite{Hohenberg:1977ym}. They characterize the behavior of
systems near a phase transition, organizing them into universality classes.
These critical exponents satisfy a number of relations among each other. One of these is the
hyperscaling relation $n \nu = 2 - \alpha$. The assumption underlying the hyperscaling relation is that all dimensionful quantities scale with their natural power of length. In particular, the entropy density should scale as a volume. In systems where the only length scale is the correlation length, this in particular
implies that the entropy density should scale like the correlation length to the $n$, where $n$ is the number of spatial dimensions. This relation can be violated
in certain cases,
and this violation is measured by the hyperscaling violation exponent $\theta$ in the following way \cite{Fisher:1986zz}
\begin{equation}\label{hypersc1}
\left ( n-\theta\right) \nu = 2 - \alpha \, ,
\end{equation}
where $\alpha$ describes the scaling of the specific heat capacity $C \sim \tau^{-\alpha}$,
while $\nu$ describes the scaling of the correlation length $\xi \sim |\tau|^{-\nu}$ near the phase transition
at temperature $T=T_c.$ $n$ are the number of spatial dimensions and $\tau$ is the reduced temperature given by $\tau = (T-T_c)/T_c$.
Such theories show conformal symmetry at the transition $T=T_c$. However,
the scaling behavior in this context is always extracted from $T\not=T_c$,
where the theory should be slightly deformed away from conformal invariance.

In mean field theory, critical exponents only depend on the form of the effective potential and so are independent of spatial dimension $n$. As such, mean field theory and hyperscaling are only consistent in a critical dimension $n_U,$ the so-called upper critical dimension. For the Ising model this critical dimension is $n_U=4$. The expectation is that for $n<n_U$ hyperscaling is valid, but mean field theory is not, whereas in $n>n_U$ mean field theory is valid and hence hyperscaling violated. As theories with a supergravity dual are intrinsically classical and often reproduce mean field results, one should presumably not be surprised that hyperscaling violation is actually the norm in these systems, as seen for example in~\cite{Buchel:2010gd,Buchel:2010ys}.
In a previous paper four of the critical coefficients were computed correctly~\cite{Maeda:2008hn}. However, the authors assume validity of the (unmodified) hyperscaling relation, i.e.~\eqref{hypersc1} with $\theta=0$. This leads to values for the two remaining exponents, which then disagree with later direct calculations of said two exponents. 
In the context of holographic bottom-up models involving charged dilatonic black holes hyperscaling has been discussed in~\cite{Gouteraux:2011ce}. An unusual scaling of the metrics discovered previously in~\cite{Charmousis:2010zz} was interpreted as a violation of the hyperscaling relation. 

So far we discussed thermal phase transitions occurring at $T_c \neq 0.$ In the present paper, we consider a phase transition at zero temperature which is therefore driven by quantum fluctuations. Such quantum phase transitions occur at a particular value $r_c$ of the control parameter $r,$ which can be a magnetic field, pressure -- or in the case of the present paper a chemical potential $\mu.$ In the case of a quantum phase transition the relations stated above hold if we replace the number of spatial dimensions, $n,$ by $n+z$. Here $z$ is the dynamic critical exponent associated with the scaling of dynamic quantities. One such quantity is the correlation time $t_\mathrm{corr} \propto \xi^z$, which represents the characteristic time scale of the dynamical system under consideration. Now, for example, the hyperscaling relation \eqref{hypersc1} with hyperscaling violating exponent $\theta$ reads
\begin{equation} \label{eq:modifiedHyperscaling}
\left( n+z - \theta \right)\nu = 2 - \alpha \, .
\end{equation}
At non-vanishing temperature, $T\neq 0$, the free energy density $f$ in the quantum critical region can be written as (see e.g.~\cite{Vojta:2003})
\begin{equation} \label{eq:fsim}
f \sim |\delta|^{2-\alpha} g\left( \frac{T}{|\delta|^{\nu z}} \right) \, ,
\end{equation}
where $\delta = r -r _c.$ The function $g$ is called scaling function and satisfies $g(T=0)=1.$

Also for quantum phase transitions a critical dimension $n_U$ exists analogous to the classical phase transitions mentioned above. But for quantum phase transitions we now have to consider the modified number of spatial dimensions. That is, we expect that for $n+z < n_U$ hyperscaling is valid, mean field theory is not. And vice versa in the case $n+z > n_U$.

Hyperscaling violation has been recently analyzed~\cite{Huijse:2011ef,Dong:2012se} holographically. In these works a scale was introduced by studying the theory at finite temperature\footnote{At small temperatures, the quantum critical region around $r=r_c$ is still governed by the critical point at $r=r_c$ and at $T_c=0$.}; in the present paper we will similarly introduce a scale via a finite chemical potential.
Neither in~\cite{Huijse:2011ef,Dong:2012se} nor in the present paper does the static coefficient $\nu$ mentioned above appear explicitly.
However, we will see that in the present paper by construction $\nu z =1.$
The assumption of hyperscaling still should mean that all dimensionful quantities have their
natural length dimension. This time the natural length scale is $\tau^{-1} \sim T^{1/z}$ if we are studying the theory at finite temperature but zero density or $\tau^{-1} \sim \mu^{1/z}$ when the chemical potential is non-zero but the temperature vanishes.
The entropy density in such a theory is given by
$s\sim\tau^{n/z}$ if hyperscaling is intact, or by $s\sim\tau^{(n-\theta)/z}$ if hyperscaling
is violated~\cite{Huijse:2011ef}. Roughly speaking, in theories with hyperscaling violation the length
dimension of a spatial volume is reduced from $n$ to $n-\theta$.

In general the gauge/gravity correspondence~\cite{Maldacena:1997re} relates quantum field theories
to gravity theories defined on spacetimes with metrics which are asymptotically
Anti-deSitter~(AdS). In~\cite{Huijse:2011ef} a particular behavior of the metric
on the gravity side was identified to be yielding non-trivial $\theta$ and $z$ on the
gauge side.
This particular behavior amounts to a specific scaling of the infrared (IR) limit of any
AdS metric
\begin{equation}
ds^2 = \rho^2 \left(-\rho^{2n(z-1)/(n-\theta)}{dt}^2
+\rho^{-2(2n-\theta)/(n-\theta)}{d\rho}^2
+{dx_i}^2\right) \, ,
\end{equation}
with the time direction $t$, $n$ spatial directions $x_i$, and the emergent
holographic direction $\rho$.
From a rescaling with constant $\lambda$ we get $t\to \lambda^{-z} t,\, x_i\to \lambda^{-1} x_i,\, \rho\to \lambda^{(n-\theta)/n} \rho$, and most prominently
$ds\to \lambda^{-\theta/n} ds$.
At finite temperature this metric will develop a horizon at $\rho=\rho_H$. The
horizon area is of size $\rho_H^n$ and is proportional to the entropy of the dual field
theory. As argued in~\cite{Huijse:2011ef}, we obtain a scaling of the entropy density
$s\sim\rho_H^n \to \lambda^{n-\theta} s$.
This correspondence was further developed in~\cite{Shaghoulian:2011aa,Dong:2012se,Hartnoll:2012wm,Singh:2012un,Narayan:2012hk,Dey:2012tg,Dey:2012rs,Perlmutter:2012he}.

In this paper we consider a quantum field theory at zero temperature, which undergoes a quantum phase transition at a critical chemical potential $\mu_c$.  Both the D3/D7-, as well as D3/D5-brane systems have a dual description in terms of a gauge theory coupled to matter fields in the fundamental representation.
In the quantum field theory these fundamental flavors have mass $M$, so at zero temperatures a chemical potential $\mu\leq \mu_c=M$ does not lead to any finite density, whereas for chemical potentials $\mu >M$ a finite charge density develops. As shown in \cite{Karch:2007br} these two phases are separated by a second order phase transition. In the holographic dual, the transition corresponds to a topology change in the embedding of the probe branes (from a configuration where the probe branes are separated from the horizon to one where they cross the horizon).
That is, in our system the chemical potential $\mu$ acts as a tunable external parameter
(analogous to the temperature in a ferromagnet). The order parameter is the charge density $d$ (analogous to the magnetization density in a ferromagnet). The mass $M$ of the fundamental flavor particles provides a microscopic scale in this system determining the critical value of $\mu$ (like the lattice constant and spin-couplings in the case of the ferromagnet). Despite the appearance of the microscopic scale $M$ our system is described by a scale invariant quantum critical theory at the 2nd order phase transition point when $\mu=\mu_c$. The non-relativistic chemical potential $\bar{\mu}\equiv\mu-\mu_c$ deforms the theory away from the critical point $\mu_c$ in the spirit of the statistical mechanics treatment reviewed above.
Our main result is that the
critical point at $\mu=\mu_c$ is governed by a scale invariant theory with non-trivial dynamical exponent $z$ and a non-zero hyperscaling violation exponent $\theta$,
which acquire the values
\begin{equation}\label{eq:exponents}
z=2\, , \qquad \theta = 1\, .
\end{equation}

This paper is structured as follows: We begin by sketching our line of argument in section~\ref{sec:argument}. This leads to our main results, the values of the scaling exponents $z$ and $\theta$.  In section~\ref{sec:embeddings} we describe our holographic model, and prove details about its thermodynamics, normal and zero sound modes in the ensuing section~\ref{sec:thermoD3D7}. A generalization to finite temperature is carried out in section~\ref{sec:finitetemp}. We discuss our results in section~\ref{sec:discussion}, while the appendix shows some details of the zero sound mode calculation.

%%%%%%%%%%%%%%%%%
\section{Scaling exponents and the main argument}
\label{sec:argument}

Consider a system with hyperscaling violation exponent $\theta$ and dynamical critical exponent $z$ in $n$ spatial dimensions. In such a system we work out scaling dimensions of thermodynamic quantities in all generality.

Let us further consider a field theory at zero temperature, which only has
one dimensionful scale, the chemical potential $\bar{\mu}$.
The chemical potential $\bar{\mu}$ can be written as the time component of a vector field and thus scales like an inverse time-coordinate with $[\bar{\mu}]=z$.
From thermodynamics at zero temperature we obtain the energy density $e $, and the grand canonical potential density
is given by $\Omega = e - \bar{\mu} d$ in terms of the energy density
$e$, charge density $d$.
 So since $[E] =  z$,  the energy density has $[e] = n-\theta +z$, because
 by definition the spatial volume scales like
$[V] = [d^nx] = -(n - \theta)$. Thus, $[e]=n-\theta+z$ and also
$[\Omega] = [e]=n-\theta+z$.
Therefore the charge density has to have
scaling dimension $[d] = n-\theta$.
Up to a sign the pressure $p$ is given by the density of the grand canonical potential and therefore
we also know $[p] = n-\theta +z$.
At zero temperature the only dimensionful scale is $\bar{\mu}$ and
we hence have to have
\begin{equation}
p=-\Omega = C_0 \bar{\mu}^{(n+z-\theta)/z}
\end{equation}
with some constant $C_0$. This implies that the energy density
is
\begin{equation} \label{eq:enr}
e = -p + \bar{\mu} d = \frac{n-\theta}{z} p \, ,
\end{equation}
where we have used $d = \partial p/\partial\bar{\mu}$.
Note that \eqref{eq:enr} is a generalization of the commonly postulated
$z e = n p $~\cite{Adams:2008wt,Kovtun:2008qy}.

Heating up this theory to finite temperature $T$ with $[T]=z$, we also obtain the scaling
behavior of the entropy density $[s]=n-\theta$.

The goal of this paper is to realize a system as it is described above.
Thus we consider a particular field theory: ${\cal N}=4$ super-Yang-Mills theory
in $3+1$ dimensions with
${\cal N}=2$ fundamental hypermultiplets, which all have the mass $M$.
These hypermultiplets can live in either $3+1$ or $2+1$ dimensions
for the cases we consider. In addition our theory
enjoys a $U(1)$-symmetry yielding a conserved charge, for which we switch on a
density $d$, thermodynamically conjugate to the chemical potential ${\mu}$.

For $\mu = M$ both systems have a second order phase transition at zero temperature. In this paper we consider the regime $\mu\approx M$ for $\mu>M$.
As an expansion parameter we use $\bar{\mu} = \mu-M \ll M$, and
consider only the leading non-vanishing contributions in $\bar{\mu}/M$.
$M$ is the rest mass of the particles in our hypermultiplets, and
the chemical potential is the energy which is needed to add one such particle to the system. Therefore the limit $\bar{\mu}\ll M$ can be thought of as a
non-relativistic limit for this theory. The non-relativistic energy-density
is given by
\begin{equation}
e=\epsilon-\epsilon_{\text{rest}} = -p +\mu d -M d = - p + \bar{\mu} d \, ,
\end{equation}
where $\epsilon$ is the relativistic energy density, $\epsilon_{\text{rest}}=M d$
the rest mass energy density of particles with mass $M$.

In this paper we consider cases in which the hypermultiplet particles can propagate
in $3+1$ and $2+1$ dimensions (corresponding to D3/D7 and D3/D5
brane setups, respectively).
Now, we state our first result about the non-relativistic energy density:
\begin{equation}\label{eq:enrp}
e=p\, , \qquad \mbox{and} \quad e = \frac{p}{2} \, ,
\end{equation}
for the $3+1$- and the $2+1$-dimensional case, respectively.
These relations will be derived from D-brane thermodynamics below.
Using \eqref{eq:enrp} in \eqref{eq:enr} immediately yields the following
relation between $\theta$ and $z$:
\begin{equation} \label{eq:scalingRelation}
3-\theta = z\, , \qquad \mbox{and} \quad 2-\theta=\frac{z}{2} \, ,
\end{equation}
for $3+1$ and $2+1$ dimensions, respectively.

Now, our second result states that the dispersion of the normal sound mode
in both the $3+1$- and $2+1$-dimensional theory is
\begin{equation}
\omega \simeq \sqrt{\frac{\bar{\mu}}{M}} \, k \, ,
\end{equation}
for $\bar{\mu}\ll M$. We will see later that this is confirmed also by the dispersion of
the zero sound mode.
The scaling dimension of the left-hand side is given by $[\omega] = z$,
whereas the right-hand side gives $[\bar{\mu}]/2 + [k] = z/2 +1.$ Therefore we conclude that $z=2$.
Using \eqref{eq:scalingRelation} we obtain for both cases
the value $\theta =1$. Thus we find non-trivial dynamical scaling $z$ and a non-zero
hyperscaling violation as claimed in the introduction in \eqref{eq:exponents}.
As we will see, mean field theory in this system yields $z=2$ and $\theta=n-2$. Hence, for the $n=3$ case our findings are consistent with mean field theory. However, in $n=2$ mean field theory would yield a vanishing hyperscaling violating exponent whereas we find again $\theta=1$.

Now let us prove case by case the two results we have used above to derive the scaling exponents in these cases.

%%%%%%%%%%%%%%%%%%%%%%%%%%%
\section{D3/D(2$n$+1) brane embeddings}
\label{sec:embeddings}

Let us now discuss the gravity dual of the field theory considered in section \ref{sec:argument}. The $\mathcal{N}=4$ supersymmetric vector multiplet is realized by $N_c$ coincident D3--branes which are aligned along the directions $0,1,2,3$ in the flat ten dimensional spacetime. The $N_f$  hypermultiplets propagating in $n+1$ dimensions (where $n$ is the number of spatial dimensions) are represented by $N_f$ D(2$n$+1)--branes~\cite{Karch:2002sh,Karch:2000gx} which are aligned as shown in figure~\ref{tab:d3D7Coords}.
Concretely, we discuss the $n=3$ D3/D7- \cite{Karch:2002sh} and the $n=2$ D3/D5-system \cite{Karch:2000gx,DeWolfe:2001pq}
in what follows.

\paragraph{Gravity setup}
\begin{figure}
\begin{center}
\begin{tabular}[c]{|c|c|c|c|c|c|c|c|c|c|c|}
\hline
 & 0 & 1 & 2 & 3 & 4 & 5 & 6 & 7 & 8 & 9 \\
 \hline
$N_c$ D3 & x & x & x & x  & & & & & & \\
 \hline
$N_f$ D7 & x & x & x & x & x & x & x & x&  & \\
  \hline
$N_f$  D5 & x & x & x &   &x &x & x & & & \\
 \hline
 \end{tabular}
\end{center}
 \caption{ \label{tab:d3D7Coords}
 Coordinate directions in which the Dp--branes extend are marked by x. While D3- and D7-branes share the four Minkowski directions, D3-- and D5--branes  share only a $2+1$-dimensional plane in the four Minkowski directions.
 }
\end{figure}

Taking the large $N_c$ limit $N_c \rightarrow \infty, \lambda = g_{YM}^2 N_c$ fixed, as well as the strong coupling limit $\lambda \gg 1$ we can replace the $N_c$ coincident D3--branes by their near-horizon geometry. In the limit $N_f \ll N_c,$ the $N_f$ D(2$n$+1) branes can be viewed as probe branes in this geometry.

The background metric which we take to be $AdS_5 \times S^5$ can be written as
\begin{equation}
ds^2 = H(r) \, \eta_{\mu\nu} \, dx^\mu dx^\nu + H^{-1}(r) \left( d\rho^2 + \rho^2 ds^2_{S^n} + dy^2 + \sum_{i=1}^{4-n} dz_i^2 \right)\, ,
\end{equation}
where $H(r) = r^2 / R^2$ with
\begin{equation}
r^2 = \rho^2 + y^2 + \sum_{i=1}^{4-n} z_i^2 \, ,
\end{equation}
and we have arranged the coordinates of $S^5$ in terms of an $S^n$ (n=2 for D3/D5 and n=3 for D3/D7) as well as coordinates $y$ and $z_i$ (where $i=1, \dots, 4-n$). These coordinates are very convenient to describe the embedding of the D(2$n$+1)--branes into $S^5.$ In particular, the D(2$n$+1)--brane wraps the $S^n,$ extends along the radial direction $\rho,$ the time direction as well as $3-n$ spatial directions. The world volume coordinates are denoted by $\xi.$ The embedding of the D(2$n$+1)--branes into $AdS_5 \times S^5$ is given by the transverse scalars $y, z_i$ and $x^a$ (where $a=n+1, \dots, 3$) as a function of the world volume coordinates $\xi$. The embeddings at hand should neither break the rotational invariance of $S^n,$ nor the Lorentz-invariance of the common directions along 0,1,2,3. Therefore the embedding is specified by $y=y(\rho)$, $z_i=z_i(\rho)$, and $x^a = x^a(\rho)$ where $a=n+1, \dots, 3.$ Here we will set $z_i(\rho)= x^a(\rho)=0.$ Besides the scalar fields, the dynamics of the probe D(2$n$+1)--branes is further
described by a $U(N_f)$-valued gauge field. Here we will restrict ourselves to the $U(1) \subset U(N_f)$ gauge group and consider a non-vanishing gauge field component $A_t(\rho).$

The dynamics of the D(2$n$+1)--brane is given by the DBI action
\begin{equation}
S_{D(2n+1)} = - N_f T_{D(2n+1)} \int d^{2n+2} \xi \sqrt{- det( g_{ab} + 2\pi \alpha^\prime F_{ab})} \, ,
\end{equation}
where $g_{ab}$ is the induced world-volume metric on the world volume of the D(2$n$+1) brane and $F_{ab}$ is the field strength tensor of the U(1) gauge field. Moreover $T_{D(2n+1)}$ is the tension of the D(2$n$+1)--brane. Since none of the fields depends on the coordinates of $S^n$ or the field theory directions in common, we can perform the integration over nine directions to get the action density, $s_{D(2n+1)}$ which is the action divided by the volume of the field theory directions in common
\begin{equation} \label{eq:actionD3Dq}
s_{D(2n+1)} \equiv \int d\rho \, \mathcal{L} =  -\mathcal{N}_n \int d\rho \rho^n \, \sqrt{1+ y'(\rho)^2
 -(2\pi \alpha')^2 A_t'(\rho)^2} \, .
\end{equation}
Here, $\mathcal{N}_n = N_f T_{D(2n+1)} V_{S^n}$ and $V_{S^n}$ is the volume of the $n$--sphere. Since the action \eqref{eq:actionD3Dq} does not depend on $y(\rho)$ and $A_t(\rho)$ explicitly we find two conserved charges
\begin{equation}\label{eq:conjugateCD}
\frac{\delta \mathcal{L}}{\delta y'(\rho)} = - c \, ,\qquad \quad
\frac{1}{(2\pi\alpha')}\frac{\delta \mathcal{L}}{\delta A_t'(\rho)} = d \, .
\end{equation}

\paragraph{General solutions at zero temperature}
Equations~\eqref{eq:conjugateCD} can be solved in terms of $y'(\rho)$ and $A_t'(\rho)$
\begin{eqnarray}\label{eq:generalyAt}
y^\prime(\rho) &=& \frac{c}{\sqrt{
{\cal N}_n^2\rho^{2n} +d^2 - c^2}}\, , \nonumber\\
A_t^\prime(\rho) &=& \frac{1}{(2\pi\alpha')} \frac{d}{\sqrt{
{\cal N}_n^2 \rho^{2n} + d^2 - c^2}} \, ,
\end{eqnarray}
which can be integrated to (written in terms of hypergeometric functions ${}_2F_1$)
\begin{eqnarray}
y(\rho) &=& \frac{c}{\sqrt{d^2-c^2}} \, \rho\,\, {}_2F_1\left(\frac{1}{2 n},\frac{1}{2};1+\frac{1}{2n};-\frac{{\cal N}_n^2 \rho^{2n}}{d^2-c^2}\right)  \, ,\\
A_t(\rho) &=& \frac{1}{(2\pi\alpha')}\frac{d}{\sqrt{d^2-c^2}} \, \rho\,\, {}_2F_1\left(\frac{1}{2 n},\frac{1}{2};1+\frac{1}{2n};-\frac{{\cal N}_n^2 \rho^{2n}}{d^2-c^2}\right) \, .
\end{eqnarray}
Note that we have chosen the integration constants such that $y(\rho =0) =0$ and $A_t(\rho=0)=0.$ The embedding functions behave near the AdS-origin (the zero temperature horizon)
$\rho \to 0$ as
\begin{equation}
y(\rho) = \frac{c}{\sqrt{d^2-c^2}} \rho + \mathcal{O}(\rho^{2 n+1}) \, ,
\end{equation}
for all D3/D(2$n$+1)-systems.

\paragraph{Boundary asymptotics}
From the asymptotic values $\lim_{\rho\to\infty} y(\rho) = M$ and
$\lim_{\rho\to\infty} A_t(\rho) = \mu/(2\pi\alpha')$ we get the relation between the constants of
motion $c,\, d$ and the physical parameters $M$ and $\mu$ for $n=3$:
\begin{equation}
c = {\cal C}_3  M \left(\mu^2- M^2\right)\, ,\qquad
d = {\cal C}_3 \mu \left(\mu^2- M^2\right) \, ,
\end{equation}
for $n=2$
\begin{equation}
c = {\cal C}_2  M \sqrt{\mu^2- M^2} \, ,\qquad
d = {\cal C}_2  \mu \sqrt{\mu^2- M^2}  \, .
\end{equation}
where we have defined
\begin{equation}
{\cal C}_3 = \frac{{\cal N}_3 \pi^{3/2}}{\Gamma(1/3)^3 \Gamma(7/6)^3} \, , \qquad\quad {\cal C}_2 = \frac{{\cal N}_2 \pi}{\Gamma(1/4)^2 \Gamma(5/4)^2} \, .
\end{equation}

\paragraph{On-shell actions}
In order to extract thermodynamic quantities, we need the on-shell action. Since
the action \eqref{eq:actionD3Dq} is divergent for $\rho \rightarrow \infty,$ we introduce a cutoff $\Lambda$ which we send to $\Lambda \rightarrow \infty$ at the end of the calculation. To obtain a finite action we add appropriate counterterms of the form
\begin{equation}
s_{\text{counter}} = \mathcal{N}_n \int d\rho \rho^n \, .
\end{equation}
Thus the renormalized on-shell action density for the D3/D7- and D3/D5-systems is given by
\begin{equation}
s_{\text{ren},n} = s_{\text{D(2$n$+1)}} + s_{\text{counter}} \, .
\end{equation}

%%%%%%%%%%%%%%%%%%%%%%%%%%%
\section{Hyperscaling violation from D3/D7 and D3/D5}
\label{sec:thermoD3D7}
In this section we briefly review the D3/D(2$n$+1)-brane setups at zero temperature, and the implications
of their stationary analytic solutions for the thermodynamics of the dual theory \cite{Karch:2007br}.
This will lead to the relations between the non-relativistic energy density $e$
and the pressure $p$.

%--------------------------------------
\subsection{Thermodynamics}
\label{sec:thermoD3D7D3D5}

\paragraph{D3/D7}
Since the density of the thermodynamical potential $\Omega$ in the grand canonical ensemble is given by the renormalized action density $s_{ren},$ we obtain
\begin{equation}
\Omega = - s_{ren}
=-\frac{\Gamma(1/3)\Gamma(7/6)}{4\sqrt{\pi}\mathcal{N}_3^{1/3}} (d^2-c^2)^{2/3}
=- \frac{1}{4}\mathcal{C}_3 (\mu^2-M^2)^2 \, .
\end{equation}
Note that the density of the thermodynamical potential $\Omega$ is (minus) the pressure $p,$
\begin{equation}
p =\frac{1}{4}\mathcal{C}_3 (\mu^2-M^2)^2 \, .
\end{equation}
In order to obtain the density of the free energy, $f,$ we Legendre transform $\Omega,$
\begin{equation}\label{eq:freeenD3D7}
f = \Omega + \mu d
=\frac{1}{4} \mathcal{C}_3 (\mu^2 - M^2) (3\mu^2 + M^2) \, .
\end{equation}
Note that the density of the free energy, $f$ is related to the (relativistic) energy density $\epsilon$ by $f=\epsilon - s T,$ where $s$ is the entropy density. Since we are working at zero temperature $f=\epsilon.$

Let us now consider a special case, where $\mu$ is only slightly larger than $M.$ Therefore we write $\mu = M + \bar{\mu}$ with $\bar{\mu} \ll M.$ In this case we get to leading order
\begin{eqnarray}\label{eq:thermoD3D7}
\Omega = - p & = & - \mathcal{C}_3 M^2 \bar{\mu}^2 \, ,\nonumber\\
\epsilon = f &=& 2 \mathcal{C}_3 M^3 \bar{\mu}
 \, ,\nonumber\\
c=d&=& 2 \mathcal{C}_3 M^2 \bar{\mu}\, .
\end{eqnarray}
Note that in particular $\epsilon \sim d M$ to leading order in $\bar{\mu}/M.$ In the following we consider $\bar{\mu}$ as a non-relativistic chemical potential. The non-relativistic energy density $e$ is then given by the difference of the relativistic energy $\epsilon$ and $dM,$ i.e. $e = \epsilon - d \, M$ where $d \, M$ should be viewed as a rest-frame energy density,
\begin{equation}
e = \epsilon - d \, M = \mathcal{C}_3 M^2 \bar{\mu}^2 \, ,
\end{equation}
to leading order in $\bar{\mu}/M.$ Note that to leading order in $\bar{\mu} / M$ we get
\begin{equation}\label{eq:epD3D7}
e = p \, ,
\end{equation}
which concludes the proof of our first result in section \ref{sec:argument}.

\paragraph{D3/D5}
Similarly we obtain
\begin{equation}
\Omega = - s_{ren}
=-\frac{\Gamma(1/4)\Gamma(5/4)}{3\sqrt{\pi \mathcal{N}_2} } (d^2-c^2)^{3/4}
=-\frac{1}{3} \mathcal{C}_2 (\mu^2-M^2)^{3/2} \, .
\end{equation}
The free energy density is then given by
\begin{equation}\label{eq:freeenD3D5}
f = \frac{1}{3} \mathcal{C}_2 \sqrt{\mu^2-M^2} (2\mu^2+M^2) \, .
\end{equation}

Considering also the limit $\mu = M + \bar{\mu}$ where $\bar{\mu} / M \ll 1$ we obtain to leading order in $\mu:$
\begin{eqnarray}\label{eq:thermoD3D5}
\Omega = - p & = & - \frac{2\sqrt{2}}{3} \mathcal{C}_2 M^{3/2} \bar{\mu}^{3/2} \, ,\nonumber\\
\epsilon = f &=& \sqrt{2} \mathcal{C}_2 M^{5/2} \bar{\mu}^{1/2} \, ,\nonumber\\
c=d&=& \sqrt{2} \mathcal{C}_2 M^{3/2} \bar{\mu}^{1/2} \, .
\end{eqnarray}
Note that as in the case of D3/D7, also $\epsilon \sim dM$ to leading order in $\bar{\mu}.$ The non-relativistic energy $e$ is given by
\begin{equation}
e = \epsilon - d \, M = \frac{\sqrt{2}}{3} \mathcal{C}_2 M^{3/2} \bar{\mu}^{3/2} \, ,
\end{equation}
to leading order in $\bar{\mu}.$ Comparing the non-relativistic energy density to the pressure we find
\begin{equation}\label{eq:epD3D5}
e = \frac{p}{2} \, ,
\end{equation}
to leading order in $\bar{\mu}/M$.

%--------------------------------------
\subsection{Normal sound}
\label{sec:sound}

Here we compute the speed of sound. Its scaling with $\bar{\mu}$ will allow
us to compute the values of exponents $\theta$ and $z$.
The speed of sound in a relativistic system is defined as
\begin{equation}
{v_s}^2 = \frac{\partial p}{\partial\epsilon} \, .
\end{equation}

From equations \eqref{eq:freeenD3D7} and \eqref{eq:thermoD3D7} we know that for the D3/D7-system we have
$p \propto (\mu^2 - M^2)^2$ and
$\epsilon \propto  (\mu^2 - M^2) (3 \mu^2 + M^2)$.
Therefore we can write
\begin{equation}\label{eq:vsD3D7}
{v_s}^2 = \frac{(\partial p / \partial \mu)}{(\partial \epsilon/\partial \mu)}
 = \frac{\mu^2 - M^2} {3 \mu^2 - M^2} \approx \frac{\bar{\mu}}{M}\, ,
 \end{equation}
 where in the last step we approximate $\bar\mu\ll M$.
For a propagating sound mode with frequency $\omega$ and momentum $k$
the scaling of the speed of sound in \eqref{eq:vsD3D7}
implies $\omega = k \sqrt{\bar{\mu}/M}+\mathcal{O}(k^2,\bar{\mu}^2)$ at small $\bar{\mu}\ll M$, and small $\omega\ll \bar\mu,\, k\ll \bar\mu$.
This concludes the proof of our second result in section \ref{sec:argument}.

Let us repeat this computation for the D3/D5 system. We obtain
\begin{equation}\label{eq:vsD3D5}
{v_s}^2 = \frac{(\partial p / \partial \mu)}{(\partial \epsilon/\partial \mu)}
=\frac{\mu^2-M^2}{2\mu^2-M^2}
\approx 2\frac{\bar{\mu}}{M}\, ,
 \end{equation}
where in the last step we again approximate $\bar\mu\ll M$.

In fact, one gets $v_s^2 \sim \bar{\mu}/M$ (and hence $z=2$) in any scale invariant non-relativistic field theory (with or without hyperscaling violation) at zero temperature. This reasoning applies whenever a linear mode exists which has a speed given by the square root of~\eqref{eq:vsD3D5}. As we have seen, in such a theory the free energy (and hence the pressure) scales as $\bar{\mu}^x$ with $x=(n+z-\theta)/z$. The value of $x$ will not be important for our argument.
In a non-relativistic theory the speed of sound is given by
\begin{equation}
v_s^2 = \frac{1}{M} \frac{\partial p}{\partial d} = \frac{\partial p}{\partial \rho} \, ,
\end{equation}
where $\rho$ is the mass density,  $\rho=d M$.
The relativistic formula reduces to the non-relativistic one when we assume that $\epsilon \approx d M$, that is the energy density is dominated by the rest mass.
Due to the scaling form of $p$ we can then calculate as above
\begin{equation}
v_s^2=\frac{1}{M} \frac{(\partial p/ \partial \bar{\mu})}{(\partial d/ \partial \bar{\mu}) } \sim \frac{\bar{\mu}}{M} \, ,
\end{equation}
for any $x$ where $x\neq 0,1.$

%--------------------------------------
\subsection{Zero sound}
\label{sec:zeroSound}

Now we make use of the dynamics of fluctuations on the D7-brane.
In particular, we find the dispersion relations of the sound mode in these fluctuations, which was identified as zero sound in \cite{Karch:2008fa}.
In our zero temperature, finite density system there is only one sound mode, which can be equivalently viewed as zero sound or normal sound.
We will confirm that the dispersion relation of the probe brane sound mode is in perfect agreement with the normal sound in the previous subsection. For the standard formulas for the speed of normal sound in terms of thermodynamic quantities to hold we basically have to assume that the low energy dynamics of the system is governed by hydrodynamics. The reason zero sounds is usually viewed as different from normal sound is that the zero $T$, finite $\mu$ regime traditionally is not identified with a hydrodynamic regime. Experience from holography however shows that for a hydrodynamic description having $\mu$ to set the scale of what constitutes low energies is completely sufficient. We find it reassuring that the explicit fluctuation calculation agrees with the prediction from thermodynamics. Here we present a very condensed version of the computation.
Details of the zero sound calculation can be found in appendix \ref{sec:zeroSoundComp}.

In order to calculate the zero sound, we consider fluctuations $\eta$ of the form
$y(t,x,\rho) = y_0(\rho) + \eta(t,x,\rho)$ around the D7-brane embedding $y_0(\rho)$.
Furthermore, we choose the following ansatz for the gauge field
\begin{eqnarray}
\mathcal{A}_t(t,x,\rho) &=& A_t(\rho) +  a_t(t,x,\rho)\, ,\\
\mathcal{A}_x(t,x,\rho) &=& a_x(t,x,\rho)\, ,\\
\mathcal{A}_\rho(t,x,\rho) &=& a_\rho(t,x,\rho)\, ,\\
\mathcal{A}_i(t,x,\rho) &=& 0 \, .
\end{eqnarray}
Let us denote
\begin{eqnarray}
f_{tx}(t,x,\rho) &=& \partial_t a_x(t,x,\rho) - \partial_x a_t(t,x,\rho)\, , \\
f_{t\rho}(t,x,\rho) &=& \partial_t a_\rho(t,x,\rho) - \partial_\rho a_t(t,x,\rho)\, , \\
f_{x\rho}(t,x,\rho &=& \partial_x a_\rho(t,x,\rho) - \partial_\rho a_x(t,x,\rho)  \, .
\end{eqnarray}

From now on, we set $R=1$ and $2\pi\alpha^\prime =1.$ Then we can write the action to second order in the fluctuations $a(t,x,\rho),$ $\eta(t,x,\rho)$ in the notation of \cite{Kulaxizi:2008kv}
\begin{eqnarray} \nonumber
S_{D(2n+1)}^{(2)} = - \frac{\mathcal{N}}{2} \int dx \int d\rho \, g(\rho)&&\left[ f_{x\rho}^2 - f_1(\rho) f_{t\rho}^2 - f_2(\rho) f_{tx}^2 - f_3(\rho) (\partial_t \eta)^2\right.\\ \nonumber
&& \ \left.+ f_4(\rho) (\partial_x \eta)^2 + f_5(\rho) (\partial_\rho \eta)^2 - 2 f_6(\rho) f_{tx} \partial_x \eta\right. \\
&& \ \left. - 2 f_7(\rho) f_{t\rho} \partial_\rho \eta \right] \, .
\end{eqnarray}
In order to generalize these expressions later, we have used the functions
$g(\rho)$ and $f_i(\rho)$. For the D3/D7 brane model these are given by
\eqref{eq:fgD3Dq} with $n=3$.

After a Fourier transformation to frequencies $\omega$, momentum $k$ (see \eqref{eq:fourier}),
change of variables from $\rho$ to $z = 1/\rho$, and the field redefinitions $X=k \eta$ and $E=k a_t + \omega a_x$, we obtain the equations of motion
\begin{eqnarray}\nonumber\label{eq:eomEX}
\ddot{E} + \left(\frac{2}{z}+ \frac{\dot{g}}{g} + \frac{1}{h f_5} \frac{k^2 \dot{f}_1}{k^2 - \omega^2 f_1} \right) \dot{E} + \frac{f_8 f_3}{z^4} (\omega^2 f_1 - k^2) E - \frac{\dot{f}_7}{h f_5} \frac{k^2 - \omega^2}{k^2 - \omega^2 f_1} \dot{X} &=& 0\, , \\ \label{eqtosolve}
\ddot{X} + \left( \frac{2}{z}+ \frac{\dot{g}}{g} + \frac{\dot{f}_5}{h f_5} \frac{k^2 - \omega^2}{k^2 - \omega^2 f_1}\right) \dot{X} + \frac{f_8 f_3}{z^4} (\omega^2 f_1 - k^2) X + \frac{\dot{f}_7}{h f_5} \frac{k^2}{k^2 - \omega^2 f_1} \dot{E} &=&0 \, .
\end{eqnarray}

We solve these equations in two distinct limits. For the first case we take the limit $z\gg1$ of
\eqref{eq:eomEX}, find the solution to those approximate equations,
and then expand that solution for $\omega z\ll 1$.
In the second case we take first the limit $\omega z\ll1$, find the solution, and expand
this solution for $z\gg 1$. Then we compare the low frequency, large z expansions of
the solutions in both cases. From this matching we obtain the dispersion relation as
\begin{equation}
\omega = \sqrt{\frac{M^2-\mu^2}{M^2-3 \mu^2}}\, k +\beta_3 k^2 +\dots \, ,
\end{equation}
which is easily expanded as $\omega \approx k \sqrt{\bar{\mu}/M}$ at small $\bar{\mu}\ll M$.
This once again concludes a proof of our second postulate in section \ref{sec:argument}.
We explicitly determine the subleading coefficient $\beta_3$ in~\eqref{eq:beta3}.

Carrying out the zero sound computation for the D3/D5 system (see appendix
\ref{sec:zeroSoundComp}), we obtain the zero sound dispersion
\begin{equation}\label{eq:dispersionD3D5}
\omega = \sqrt{\frac{\mu^2-M^2}{2\mu^2 - M^2}}\, k +\beta_2 k^2 +\dots \, ,
\end{equation}
which again can be expanded as $\omega \approx k \sqrt{2 \bar{\mu}/M}$ at small $\bar{\mu}\ll M$. This confirms the scaling found from our computation of the normal sound
speed~\eqref{eq:vsD3D5}.
Again, we  explicitly determine the subleading coefficient $\beta_2$ in~\eqref{eq:beta2}.

This completes the argument made in section \ref{sec:argument}.
The sound dispersion relations fix $z=2$ and then thermodynamics for both $n=3$ and $n=2$ gives
\begin{equation}
z = 2 \, , \qquad \theta = 1 \, ,
\end{equation}
which shows that hyperscaling is violated in the D3/D7- and D3/D5-probe brane systems near their phase transition. This is the main result of the present paper and we return to discussing it in section~\ref{sec:discussion}.

%%%%%%%%%%%%%%%%%
\section{Generalization to finite temperature}
\label{sec:finitetemp}
%%%%%%%%%%%%%%%%%

In this section we turn on a small temperature $T \ll \bar{\mu}.$ In order to derive the free energy we have to embed the massive D(2$n$+1)-brane into the background of $N_c$ black 3-branes.
Such massive D(2$n$+1)-brane embeddings at non-zero temperature and chemical potential are not known analytically. The blackening factors in the metric lead to more complicated differential equations for these embeddings compared to the previous case.

Fortunately, for small temperatures $T \ll \bar{\mu}$ we can bypass the whole calculation. According to \cite{Karch:2009eb} the correction of the free energy density to first order in the temperature $T$ is given by
\begin{equation}
f(\mu, M, T) = f(\mu, M,T=0) + \pi d T + \mathcal{O}(T^2) \, .
\end{equation}
where $f(\bar{\mu}, T=0)$ is the free energy of the relativistic system at zero temperature given by equations~\eqref{eq:freeenD3D7} and~\eqref{eq:freeenD3D5} in the case of D7-branes and D5-branes, respectively. In order to obtain the non-relativistic result for the free energy we have to subtract the mass density $dM,$ i.e.
\begin{equation}
f_{non-rel.}(\mu,M,T) = f(\mu,M,T) - d M \, .
\end{equation}

In the case of D7-branes, the free energy to first order in the temperature is given by
\begin{equation}
f_{non-rel.}(\mu,M,T) = \frac{1}{4} \mathcal{C}_3 (\mu^2 - M^2) (3\mu^2 + M^2 + 4\pi \mu T- 4 M \mu) + \mathcal{O}(T^2)\, ,
\end{equation}
and in particular for $\mu = M + \bar{\mu}$ to leading order in $\bar{\mu}/M,$
\begin{equation}
f_{non-rel.}(\bar{\mu}, M, T) = \mathcal{C}_3 M^2 \bar{\mu}^2 \Big[ 1 + \frac{2\pi T}{\bar{\mu}} \Big] \, .
\end{equation}
We see that $f_{non-rel.}(\bar{\mu}, T)$ can be written in the form
\begin{equation}
f_{non-rel.}(\bar{\mu}, T) \sim \bar{\mu}^2 g\left(\frac{T}{\bar{\mu}}\right) \, ,
\end{equation}
where $g(0)=1.$ In the analysis above we have determined the first coefficient in a Taylor expansion of the function $g.$ Comparing with~\eqref{eq:fsim} we conclude that
\begin{equation}
z \nu =1 \, , \qquad\quad \alpha = 0 \, .
\end{equation}
Since $z=2$ we conclude $\nu = 1/2.$ Together with $\theta = 1$ and $n=3,$ we see that the modified hyperscaling relation
\begin{equation}
(n+z-\theta)\nu = 2-\alpha \, .
\end{equation}

Let us repeat the analysis for the D3/D5 system. Then the free energy density to first order in the temperature reads
\begin{equation}
f_{non-rel.}(\bar{\mu}, T) = \frac{\sqrt{2}}{3} \mathcal{C}_2 M^{3/2} \bar{\mu}^{3/2} \left( 1 + 3\pi \frac{T}{\bar{\mu}} + \mathcal{O}\left(\frac{T}{\bar{\mu}}\right) \right) \, .
\end{equation}
In this case we see that
\begin{equation}
z \nu=1 \, , \qquad\quad \alpha = 1/2 \, .
\end{equation}
Note also that the assignment $\alpha = 1/2 = \nu,$ $z=2, n=2$ and $\theta = 1$ satisfies the modified hyperscaling relation.

%%%%%%%%%%%%%%%%%
\section{Discussion}
\label{sec:discussion}

In this paper we have shown that the neighborhood of the second order phase transition governing the onset of finite density in the D3/D(2$n$+1)-systems is described by a non-relativistic, scale invariant field theory, which exhibits hyperscaling violation. The fixed point theory governs the non-relativistic limit where the chemical potential is just barely above the mass $M$. The quantities that scale non-trivially are the non-relativistic energy density (that is the relativistic energy density minus the rest mass of all the particles) as well as the non-relativistic chemical potential $\bar{\mu} = \mu -M$. It is the appearance of this extra scale $M$ that allows the free energy to scale differently than what one would expect from naive dimensional analysis. In both holographic systems we studied, one with $n=3$ and one with $n=2$ spatial dimensions, we find $\theta=1$ and $z=2$.
To put our results into context, we should compare and contrast our strong coupling results with the same systems at weak coupling. At weak coupling, one would expect a finite chemical potential to drive the scalars into a condensed phase. The weakly coupled system should be well described by mean field theory. That is, we simply need to study the value of the effective potential at its minimum. Writing, following Landau, the most general potential for the scalar condensate $|X|^2$ consistent with symmetries one finds (in any spatial dimension $n$):
\begin{equation}
V=(M^2-\mu^2) |X|^2 + g |X| ^4 + \ldots \sim  -\bar{\mu} X^2 +  g X^4 + \ldots \, .
\end{equation}
Solving for the condensate one finds $|X|^2 \sim (\mu^2-M^2) \sim \bar{\mu}$ and hence the density of the grand canonical potential at the minimum scales as
\begin{equation}
p \sim (\mu^2-M^2)^2 \sim \bar{\mu}^2\, .
\end{equation}
We have shown that an equation of state of the form $p \sim \bar{\mu}^x$ always gives rise to a sound mode with dynamical critical exponent $z=2$. Since we argued that $p$ scales as $ p \sim \bar{\mu}^{\frac{n+z-\theta}{z}}$ we see that $p \sim {\bar{\mu}}^2$ then fixes the hyperscaling violating exponent in mean field theory to $\theta=n-z$.
This mean field scaling is exactly what we find in the D3/D7 system with $n=3$.
For the D3/D5 system however mean field theory would predict $\theta=0$ while the brane system realizes $\theta=1$.
We find that a generalization of our scaling limit to small temperatures yields critical exponents $\alpha=0$ and $\nu=1/2$ for the field theory dual to our D3/D7-system, as well as $\alpha=1/2$ and $\nu=1/2$ for D3/D5. With these values the modified hyperscaling relation~\eqref{eq:modifiedHyperscaling} is satisfied for both of these theories.

Perhaps even more interesting is the fact that for the D3/D5 system at zero temperature we have $\theta=n-1$ which, according to the analysis of \cite{Huijse:2011ef} is exactly the value at which the system exhibits a logarithmic enhancement of the entanglement entropy as first pointed out in \cite{Ogawa:2011bz}. This behavior was seen as evidence for the existence of a Fermi surface in this system. It would be very interesting to see what the entanglement entropy is for the D3/D5 system. Unfortunately in order to use the Ryu-Takayanagi prescription \cite{Ryu:2006bv} one seems to require the fully backreacted solution, so such a calculation is not currently feasible. Curiously, if we were to allow a potential of the form $V\sim -\bar{\mu} X^2 + g_6 X^6$, then the mean field theory prediction would be $\theta=n-1$, consistent with our D3/D5 result.

Our D3/D5-brane system realizes a field theory in its critical dimension, i.e. $n=2$. Recently, a lattice computation of $\phi^4$-theory at finite chemical potential in $n=3$ dimensions has been performed~\cite{Gattringer:2012df}. We suggest to perform a similar lattice computation of the system in its critical dimension, namely 2 spatial dimensions, $n=2$.
It would then be very interesting to qualitatively compare our critical scaling results for the D3/D5-brane system in its critical dimension to such a lattice computation of $\phi^4$-theory in its critical dimension. Apart from bosonic matter the field theory dual to our D3/D5-brane system also contains fermions, whose influence on the critical exponents is not clear at this point.

In this paper we have not considered the third D3/D(2$n$+1)-brane system with 4 ND directions, namely D3/D3 \cite{Constable:2002xt}. Complications arise in this case from the fact that the boundary expansions of fields contain logarithms, see for example~\cite{Marolf:2006nd,Jensen:2010em,Gao:2012yw}. We will consider this case in a separate work, but already collect our zero sound results for D3/D3 in the appendix.

The non-relativistic fixed points whose scaling behavior we have uncovered should clearly be further investigated. One could for example switch on external magnetic fields. The presence of a second scale would allow for non-trivial tests of our proposal. We have only analyzed in detail the 3 simplest probe brane system based on an AdS$_5$ $\times$ $S^5$ background. It would be very interesting to see if the general Dp/Dq system exhibits a similar behavior. In that case already the strongly coupled gauge theory itself violates hyperscaling \cite{Dong:2012se}. So it would be very interesting to see what, if any, scaling is preserved at the non-relativistic fixed point.

%%%%%%%%%%%%%%%%%
\acknowledgments
We thank Stefan Janiszewski, Per Kraus and Eric Perlmutter for helpful discussions. MA was supported by the National Science Foundation under Grant No. NSF PHY-07-57702. MK and AK are currently supported by the US Department of Energy under contract number DE-FGO2-96ER40956.

%%%%%%%%%%%%%%%%%
\begin{appendix}
\section{Zero sound of D3/D7, D3/D5, and D3/D3}
\label{sec:zeroSoundComp}
Here we explicitly carry out the computation of zero sound
modes in the D3/D(2$n$+1) systems studied in the body of this paper. Note that
we follow in large parts the methods presented in \cite{Kulaxizi:2008kv,Ammon:2011hz}.

In general the functions $g(\rho)$ and $f_i(\rho)$ for a D3/D(2$n$+1)-system with $n=1,\,2,\,3$
are given by
\begin{equation}
\begin{split}\label{eq:fgD3Dq}
g(\rho) = \frac{\rho^n}{\sqrt{1+y_0^\prime(\rho)^2 - A_t^\prime(\rho)^2}} \, ,
\qquad
 f_1(\rho) =\frac{1+y_0^\prime(\rho)^2}{1+y_0^\prime(\rho)^2 - A_t^\prime(\rho)^2} \, , \\
f_2(\rho) = \frac{1+y_0^\prime(\rho)^2}{(\rho^2 + y_0(\rho)^2)^2} \, ,
\qquad
f_3(\rho) = \frac{1}{(\rho^2 + y_0(\rho)^2)^2} \, ,\\
f_4(\rho) =  \frac{1-A_t^\prime(\rho)^2}{(\rho^2 + y_0(\rho)^2)^2}\, ,
\qquad
f_5(\rho) =\frac{1-A_t^\prime(\rho)^2}{1+y_0^\prime(\rho)^2 - A_t^\prime(\rho)^2}\, , \\
f_6(\rho) = \frac{y_0^\prime(\rho) A_t^\prime(\rho)}{(\rho^2 + y_0(\rho)^2)^2} \, ,
\qquad
f_7(\rho) =\frac{y_0^\prime(\rho) A_t^\prime(\rho)}{1+y_0^\prime(\rho)^2 - A_t^\prime(\rho)^2}\, ,\\
f_8(\rho) = \frac{\rho^{2n}}{\rho^{2n} + d^2 - c^2} \, ,
\end{split}
\end{equation}
The general background solutions are given by \eqref{eq:generalyAt}.

Now we show how to obtain the fluctuation equations of motion in the form \eqref{eq:eomEX}.
In order to express the field equations in a convenient way, let us perform a Fourier transformation to momentum space
\begin{eqnarray}\label{eq:fourier}
a_M(t,x,\rho) &=& \int \frac{d^{n+1} k}{(2\pi)^{n+1}} \, e^{i k_\mu x^\mu} a_M(k^\mu, \rho) \, , \\
\eta(t,x,\rho) &=& \int \frac{d^{n+1} k}{(2\pi)^{n+1}} \, e^{i k_\mu x^\mu} \eta(k^\mu, \rho) \, .
\end{eqnarray}
Using rotational invariance of the $n$ spatial directions, we can consider without loss of generality $k_\mu = (-\omega, k,0,\dots,0).$ The equations of motion in the gauge $a_\rho=0$ read
\begin{eqnarray}\nonumber
\partial_\rho[ g (f_1 \partial_\rho a_t - f_7 \partial_\rho \eta) ] - \omega \, k g f_2 a_x - k^2 g f_2 a_t + k^2 g f_6 \eta &=&0 \, ,\\ \label{eq:DGL1}
\partial_\rho[ g \partial_\rho a_x ] + \omega k g f_2 a_t + \omega^2 g f_2 a_x - \omega k g f_6 \eta &=& 0 \, , \\ \nonumber
\partial_\rho[g (f_5 \partial_\rho \eta + f_7 \partial_\rho a_t) ] + \frac{\omega^2 g \eta}{(\rho^2+y_0(\rho)^2)^2} - k^2 g f_4 \eta - k \omega g f_6 a_x - k^2 g f_6 a_t &=&0  \, .
\end{eqnarray}
We also have to impose the Gauss constraint given by
\begin{equation}
k a_x^\prime + \omega f_1 a_t^\prime - \omega f_7 \eta^\prime =0 \, .
\end{equation}
Using the gauge invariant combination
\begin{equation}
E = k a_t + \omega a_x \, ,
\end{equation}
as well as the Gauss constraint, we can express $a_t^\prime$ and $a_x^\prime$ in terms of
\begin{eqnarray*}
a_t^\prime &=& \frac{k}{k^2 - \omega^2 f_1} \left( E^\prime - \frac{\omega^2}{k} f_7 \eta^\prime \right) \, ,\\
a_x^\prime &=& - \frac{\omega f_1}{k^2 - \omega^2 f_1} E^\prime + \frac{k\omega}{k^2 - \omega^2 f_1}f_7 \eta^\prime\, .
\end{eqnarray*}
Using the Gauss constraint, the first and second line of \eqref{eq:DGL1} agree. Therefore we are left with two differential equations which look like
\begin{eqnarray}
&&E'' + \left( \frac{g'}{g} + \frac{f_1'}{f_1} + \frac{\omega^2 f_1'}{k^2 - \omega^2 f_1} \right) E' + f_8 f_3 (\omega^2 f_1 - k^2) E -\\ \nonumber
&&-k \frac{f_7}{f_1} \left[ \eta'' + \left( \frac{g'}{g} + \frac{f_7'}{f_7} + \frac{\omega^2 f_1'}{k^2 - \omega^2 f_1} \right) \eta' + f_8 f_3 (\omega^2 f_1 - k^2) \eta \right] =0\, , \\ \nonumber
&&\frac{f_7}{f_5 k} \frac{k^2}{k^2 - h \omega^2} \left[ E'' + \left( \frac{g'}{g} + \frac{f_7'}{f_7} + \frac{\omega^2 f_1'}{k^2 - \omega^2 f_1} \right)  E'  + f_8 f_3 (\omega^2 f_1 - k^2) E \right] + \\ \nonumber
&& + \eta'' + \left(  \frac{g'}{g} + \frac{f_5'}{f_5} + \frac{\omega^2 f_1'}{k^2 - \omega^2 f_1} - \frac{\omega^2 h'}{k^2 - \omega^2 h} \right) \eta' + f_8 f_3 (\omega^2 f_1 - k^2) \eta =0 \, .
\end{eqnarray}
Here, we have used
\begin{equation}
h = \frac{1}{f_4 (\rho^2+y(\rho)^2)^2} = \frac{f_7^2 + f_1 f_5}{f_5} = \frac{\rho^{2n} + d^2 - c^2}{\rho^{2n} - c^2} \, .
\end{equation}

Let us introduce
$$ X = k \eta\, ,$$
and change variables from $\rho$ to $z,$ given by $z = 1/\rho.$ The equations of motion read
\begin{eqnarray}
&&\ddot{E} + \left(\frac{2}{z}+ \frac{\dot{g}}{g} + \frac{\dot{f}_1}{f_1} + \frac{\omega^2 \dot{f}_1}{k^2 - \omega^2 f_1} \right) \dot{E} + \frac{f_8 f_3}{z^4} (\omega^2 f_1 - k^2) E -\\ \nonumber
&&-\frac{f_7}{f_1} \left[ \ddot{X} + \left( \frac{2}{z}+\frac{\dot{g}}{g} + \frac{\dot{f}_7}{f_7} + \frac{\omega^2 \dot{f}_1}{k^2 - \omega^2 f_1} \right) \dot{X} + \frac{f_8 f_3}{z^4} (\omega^2 f_1 - k^2) X \right] =0 \, ,\\ \nonumber
&&\frac{f_7}{f_5} \frac{k^2}{k^2 - h \omega^2} \left[ \ddot{E} + \left(\frac{2}{z}+ \frac{\dot{g}}{g} + \frac{\dot{f}_7}{f_7} + \frac{\omega^2 \dot{f}_1}{k^2 - \omega^2 f_1} \right)  \dot{E}  + \frac{f_8 f_3}{z^4} (\omega^2 f_1 - k^2) E \right] + \\ \nonumber
&& + \ddot{X} + \left( \frac{2}{z}+ \frac{\dot{g}}{g} + \frac{\dot{f}_5}{f_5} + \frac{\omega^2 \dot{f}_1}{k^2 - \omega^2 f_1} - \frac{\omega^2 \dot{h}}{k^2 - \omega^2 h} \right) \dot{X} + \frac{f_8 f_3}{z^4} (\omega^2 f_1 - k^2) X =0 \, .
\end{eqnarray}
By adding and subtracting the equations of motion (with adequate prefactors) we can rewrite the equations of motion in the form given in \eqref{eq:eomEX}.

We solve the two equations \eqref{eq:eomEX} in two different limits.
First we consider the near-horizon limit $z \gg 1$.
Using the precise form of $f_i,$ we can write the equations
for the cases $n=3$, $n=2$ and $n=1$ (D3/D7, D3/D5 and D3/D3 systems respectively)
in the form
\begin{eqnarray}
\ddot{E}(z) + \frac{2}{z} \dot{E}(z) + \omega^2 \frac{d^2 - c^2}{d^2} E(x) &=&0 \, , \\
\ddot{X}(z) + \frac{2}{z} \dot{X}(z) + \omega^2 \frac{d^2 - c^2}{d^2} X(x) &=&0 \, .
\end{eqnarray}
Note that the equations of motion are not coupled in the near-horizon limit. Using
$$\Omega = \omega \sqrt{\frac{d^2-c^2}{d^2}}\, ,$$
we can write the solution of these two decoupled equations in the form
\begin{equation}
E(z) = A \frac{e^{i \Omega z}}{z} \, , \quad \qquad X(z) = B \frac{e^{i \Omega z}}{z} \, .
\end{equation}
In addition, if we consider $z$ in the limit $\Omega z \ll 1$
\begin{equation}\label{eq:firstLimitEX}
E(z) = \frac{A}{z} + i \Omega A + \dots \, , \quad \qquad X(z) = \frac{B}{z} + i \Omega B + \dots\, .
\end{equation}

Now consider \eqref{eqtosolve} for all $z$ but in the low frequency limit. In this limit we can drop the terms proportional to $E$ and $X.$ Therefore the equations read
\begin{eqnarray}\nonumber
\ddot{E} + \left(\frac{2}{z}+ \frac{\dot{g}}{g} + \frac{1}{h f_5} \frac{k^2 \dot{f}_1}{k^2 - \omega^2 f_1} \right) \dot{E} - \frac{\dot{f}_7}{h f_5} \frac{k^2 - \omega^2}{k^2 - \omega^2 f_1} \dot{X} &=& 0 \, ,\\ \nonumber
\ddot{X} + \left( \frac{2}{z}+ \frac{\dot{g}}{g} + \frac{\dot{f}_5}{h f_5} \frac{k^2 - \omega^2}{k^2 - \omega^2 f_1}\right) \dot{X} + \frac{\dot{f}_7}{h f_5} \frac{k^2}{k^2 - \omega^2 f_1} \dot{E} &=&0 \, .
\end{eqnarray}

Let us now solve these equations of motion. It is convenient to write $E(z)$ and $X(z)$ in the form
\begin{eqnarray}
\dot{E}(z) = \frac{1}{g(z)} \frac{F(z)}{z^2 (1+(d^2 - c^2) z^{2n})} = \frac{z^{n-2} F(z)}{(1+(d^2 - c^2) z^{2n})^{3/2}} \, ,\\
\dot{X}(z) = \frac{1}{g(z)} \frac{G(z)}{z^2 (1+(d^2 - c^2) z^{2n})} = \frac{z^{n-2} G(z)}{(1+(d^2 - c^2) z^{2n})^{3/2}}\, .
\end{eqnarray}

Then we obtain
\begin{equation}
\ddot{E}(z) = \frac{1}{g(z)} \frac{F(z)}{z^2 (1+(d^2 - c^2) z^{2n})} \left[ - \frac{\dot{g}}{g} - \frac{2}{z} + \frac{\dot{F}(z)}{F(z)} - \frac{2n (d^2-c^2) z^{2n-1}}{1+(d^2 - c^2) z^{2n})}    \right] \, ,
\end{equation}
as well as a similar expression for $\ddot{X}(z)$ where we only have to exchange $F$ by $G.$ The equations of motion read
\begin{eqnarray}
\dot{F} + \left( \frac{\dot{f}_1}{h f_5} \frac{k^2}{k^2 - \omega^2 f_1} - \frac{2n (d^2-c^2) z^{2n-1}}{1+(d^2-c^2) z^{2n}} \right) F - \frac{\dot{f}_7}{h f_5} \frac{k^2 - \omega^2}{k^2 - \omega^2 f_1} G(z) &=& 0 \, ,\\
\dot{G} + \left( \frac{\dot{f}_5}{h f_5} \frac{k^2 - \omega^2}{k^2 - \omega^2 f_1} - \frac{2n (d^2-c^2) z^{2n-1}}{1+(d^2-c^2) z^{2n}} \right) G + \frac{\dot{f}_7}{h f_5} \frac{k^2}{k^2 - \omega^2 f_1} G(z) &=& 0 \, . \\
\end{eqnarray}

After some rewriting, we see that these equations can be solved by the Ansatz
\begin{eqnarray}
F(z) &=& D_1 + D_2 z^{2n} \, , \\
G(z) &=& C_1 + C_2 z^{2n} \, ,
\end{eqnarray}
we can determine $D_1$ and $D_2$ as a function of $C_1, C_2, \omega, k$ and $c,d.$ We obtain
\begin{eqnarray}\label{eq:D1D2}
D_1 &=& \frac{(C_1 d^2-C_2)(k^2-\omega^2)}{c d k^2}\, ,\nonumber\\
D_2 &=& \frac{c^2 k^2 C_2+(c^2-d^2) (C_1 d^2-C_2) \omega^2}{c d k^2}\, .
\end{eqnarray}
Note that these two equation hold for all three cases $n=3,2,1$.
Then we have
\begin{equation}
\begin{split}
F(z)=\frac{1}{c d k^2}\Bigg[ {C_2} \left(k^2 \left(-1+c^2 z^{2 n}\right)+\left(1+\left(-c^2+d^2\right) z^{2 n}\right) \omega ^2\right)\\
+C_1 \left(d^2 k^2-d^2 \left(1+\left(-c^2+d^2\right) z^{2n}\right) \omega ^2\right)\Bigg ] \, .
\end{split}
\end{equation}
Therefore $E(z)$ and $X(z)$ are given by
\begin{eqnarray}\label{int}
E(z) &=& C_0 + \int_0^z \frac{dx \, x^{n-2} F(x)}{(1+(d^2 - c^2) x^{2n})^{3/2}} \, , \nonumber \\
X(z) &=& D_0 + \int_0^z \frac{dx \, x^{n-2} G(x)}{(1+(d^2 - c^2) x^{2n})^{3/2}}\, .
\end{eqnarray}
Note that for $z \rightarrow 0,$ $E(z)$ and $X(z)$ behave as
\begin{equation}
E(z) = C_0 + \mathcal{O}(z^{n-1})\, , \quad X(z) = D_0 + \mathcal{O}(z^{n-1}) \quad \text{for} \, n=3,2\, ,
\end{equation}
and in the case $n=1$, we have
\begin{equation}
E(z) = C_0 + C_1 \log z+\dots\, , \quad X(z) = D_0 + D_1 \log z+\dots \quad \text{for} \, n=1\, .
\end{equation}
Since we want to determine the dispersion relation of the lowest-lying mode, we can set the sources $C_0$ and $D_0$ to zero, i.e. $C_0 = D_0 =0$ for the cases $n=3,\,2$. In contrast to that for D3/D3 with $n=1$ the sources are not identified with the constant terms
but with the coefficients in front of the logarithm, see also~\cite{Marolf:2006nd,Jensen:2010em,Gao:2012yw}. So for D3/D3 we have
$D_1=0=C_1$. Comparing this to \eqref{eq:D1D2} we immediately learn that our mode has to have the dispersion relation
\begin{equation}
\omega=k \qquad \text{for}\,\, n=1\, ,
\end{equation}
implying $D_2/C_2=d/c$. Note that this is in agreement with \eqref{eq:firstLimitEX}
if $A$, $B$, $C_0$ and $D_0$ are chosen accordingly.

Now let us perform the integration in equation \eqref{int} -- note that $F$ and $G$ are given by the constants $C_1$ and $C_2.$ For large $z,$ in the
cases $n=3,\,2$ we obtain
\begin{eqnarray}
E &=& C_0 + b_1 C_1 + b_2 C_2 + \frac{a_1 C_1}{z} + \frac{a_2 C_2}{z} +\dots\, ,\\
X &=& D_0 + \tilde{b}_1 C_1 + \tilde{b}_2 C_2 + \frac{\tilde{a}_1 C_1}{z} + \frac{\tilde{a}_2 C_2}{z} +\dots\, ,
\end{eqnarray}
while for $n=1$, we get
\begin{eqnarray}
E &=& C_0 -\frac{D_2}{z} (d^2-c^2)^{-3/2} +\dots \, ,\\
X &=& D_0 -\frac{C_2}{z} (d^2-c^2)^{-3/2}+\dots \, .
\end{eqnarray}
We obtain all the
$a_1, a_2, b_1, b_2$ and $\tilde{a}_1, \tilde{a}_2, \tilde{b}_1, \tilde{b}_2$ explicitly, but refrain from listing them here.

Now let us set $C_0=D_0=0$ and let us match these solutions to the near-horizon solution in the cases $n=3,\,2$, i.e.
\begin{eqnarray}
b_1 C_1 + b_2 C_2 + \frac{a_1 C_1}{z} + \frac{a_2 C_2}{z} &=& \frac{A}{z} + i \Omega A \, ,\\
\tilde{b}_1 C_1 + \tilde{b}_2 C_2 + \frac{\tilde{a}_1 C_1}{z} + \frac{\tilde{a}_2 C_2}{z} &=& \frac{B}{z} + i \Omega B \, .
\end{eqnarray}
Therefore
\begin{eqnarray}
b_1 C_1 + b_2 C_2 &=& i \Omega A \, ,\\
a_1 C_1 + a_2 C_2 &=& A \, ,\\
\tilde{b}_1 C_1 + \tilde{b}_2 C_2 &=& i \Omega B \, ,\\
\tilde{a}_1 C_1 + \tilde{a}_2 C_2 &=& B\, ,
\end{eqnarray}
which means
\begin{eqnarray}
b_1 C_1 + b_2 C_2 - i \Omega (a_1 C_1 + a_2 C_2) &=& 0 \, , \\
\tilde{b}_1 C_1 + \tilde{b}_2 C_2 - i \Omega (\tilde{a}_1 C_1 + \tilde{a}_2 C_2) &=& 0 \, .
\end{eqnarray}
In order to have a non-trivial solution for $C_1, C_2$, the determinant of the matrix has to vanish. Inserting now $\omega = \alpha k + \beta_n k^2 + \dots$ we can determine $\alpha$ and $\beta_n.$ The latter Ansatz for $\omega$ is justified if we
assume an analytic dispersion relation, non-integer powers of $k$ would
violate this assumption.

For D3/D(2$n$+1) we finally get the dispersion relations
\begin{equation}
\omega = \sqrt{\frac{M^2-\mu^2}{M^2-n \mu^2}}\, k +\beta_n k^2 \, ,
\end{equation}
where the subleading terms are given for $n=3$ by
\begin{equation}\label{eq:beta3}
\beta_3=-\frac{3 i \sqrt{\pi}}{\Gamma(\frac{1}{6})  \Gamma(\frac{4}{3})} \frac{d (d^2-c^2)^{4/3}}{(3d^2-c^2)^2} \, ,
\end{equation}
and for $n=2$ by
\begin{equation}\label{eq:beta2}
\beta_2=-\frac{4i\sqrt{\pi}}{ \Gamma(\frac{1}{4})^2} \frac{d (d^2-c^2)^{5/4}}{(2d^2-c^2)^2} \, .
\end{equation}
Note that our result for $n=3$ agrees with the result previously obtained for this case in~\cite{Davison:2011ek} (equation (30) therein).

\end{appendix}

%%%%%%%%%%%%%%%%%
%\bibliographystyle{JHEP}
%\bibliography{bibliography}
\providecommand{\href}[2]{#2}\begingroup\raggedright\endgroup

\end{document}